\documentclass{SciPost}

\hypersetup{
    colorlinks,
    linkcolor={red!50!black},
    citecolor={blue!50!black},
    urlcolor={blue!80!black}
}

\usepackage[caption=false]{subfig}
\DeclareSymbolFont{usualmathcal}{OMS}{cmsy}{m}{n}
\DeclareSymbolFontAlphabet{\mathcal}{usualmathcal}

\fancypagestyle{SPstyle}{
\fancyhf{}
\lhead{\colorbox{scipostblue}{\bf \color{white} ~SciPost Physics }}
\rhead{{\bf \color{scipostdeepblue} ~Submission }}

\fancyfoot[C]{\textbf{\thepage}}
}

\usepackage{hyperref}

\usepackage{amsmath}
\usepackage{amssymb}
\usepackage{amsthm}

\usepackage{dsfont} 
\usepackage{graphicx}

\usepackage[normalem]{ulem}

\usepackage{tikz}

\allowdisplaybreaks

\newcommand{\bra}[1]{\mbox{$\langle \, {#1}\, |$}}
\newcommand{\ket}[1]{\mbox{$| \, {#1}\, \rangle$}}

\newcommand{\bfk}{\boldsymbol{k}}

\begin{document}

\pagestyle{SPstyle}

\begin{center}{\Large \textbf{\color{scipostdeepblue}{
Emergent Hydrodynamics in an Exclusion Process with Long-Range Interactions
\\
}}}\end{center}

\begin{center}\textbf{
A. Zahra\textsuperscript{1,2$\dagger$},
J. Dubail\textsuperscript{1,3}, and 
G. M. Sch\"utz\textsuperscript{2$\star$} 
}\end{center}

\begin{center}
{\bf 1} Laboratoire de Physique et Chimie Th\'eoriques, University of Lorraine, Nancy, France\\
{\bf 2} Centro de An\'alise Matem\'atica, Geometria e Sistemas Din\^amicos, Departamento de Matem\'atica, Instituto Superior T\'ecnico, Universidade de Lisboa, Lisbon, Portugal\\
{\bf 3} Centre Europ\'een de Sciences Quantiques and ISIS (UMR 7006), University of Strasbourg and CNRS, Strasbourg, France
\\[\baselineskip]
$\dagger$ \href{mailto:ali.zahra@univ-lorraine.fr}{\small ali.zahra@univ-lorraine.fr},
\quad
$\star$ \href{mailto:gunter.schuetz@tecnico.ulisboa.pt}{\small gunter.schuetz@tecnico.ulisboa.pt}
\end{center}

\section*{\color{scipostdeepblue}{Abstract}}
\textbf{\boldmath{%
We study the \emph{symmetric Dyson exclusion process} (SDEP)---a lattice gas with exclusion and long-range, Coulomb–type interactions that emerge both as the maximal-activity limit of the symmetric exclusion process and as a discrete version of Dyson's Brownian motion on the unitary group. Exploiting an exact ground-state (Doob) transform, we map the stochastic generator of the SDEP onto the spin-$\tfrac12$ XX quantum chain, which in turn admits a free-fermion representation. 
At macroscopic scales we conjecture that the SDEP displays \emph{ballistic} (Eulerian) scaling  non-local hydrodynamics governed by the equation
$$
\partial_t \rho+\partial_x j[\rho]=0,\qquad
j[\rho](x,t)=\frac{1}{\pi}\sin\!\bigl(\pi\rho(x,t)\bigr)\,\sinh\!\bigl(\pi\mathcal{H}\rho(x,t)\bigr),
$$
where \(\mathcal{H}\) is the Hilbert transform, making the current a genuinely non-local functional of the density.  This non-local one-field description is equivalent to a local two-field “complex Hopf’’ system for finite particle density. Closed evolution formulas allow us to solve the melting of single- and double-block initial states, producing limit shapes and arctic curves that agree with large-scale Monte-Carlo simulations. The model thus offers a tractable example of emergent non-local hydrodynamics driven by long-range interactions.
}}

\vspace{\baselineskip}




\vspace{10pt}


\section{Introduction}

Stochastic interacting particle systems  \cite{Spoh91,Kipn99,Ligg10} are classical Markov processes defined on a lattice where particles jump with rates that depend on the position of other particles. Ergodic particle systems with one 
conserved species of particles on a ring with $L$ sites have for any fixed 
number $N$ of particles a unique stationary 
distribution. Because of the conservation of particles,
the expected number $\rho_k(t)$ of particles on a site $k$ at time
$t$ satisfies for any such 
process the discrete continuity equation
\begin{equation}
\frac{\rm d}{\rm dt} \rho_k(t) = j_{k-1}(t) - j_{k}(t)
\label{dcegen}
\end{equation}
where the expectation is taken w.r.t. the distribution of the
particle configuration at microscopic time $t$, starting at $t=0$ from 
some given initial distribution. The current $j_{k}(t)$ is the average
net number of particles that cross the lattice bond $(k,k+1)$ in an 
infinitesimal time  interval $ \mathrm{d} t$. It is given by the expectation 
of the so-called instantaneous current $j^{{\rm inst}}_{k}$
which is a random number that is given by the action of the infinitesimal generator of the Markov process
on the (random) particle number $n_k(t)$ at time $t$.

The discrete continuity equation \eqref{dcegen}, even though mathematically exact, 
does not in general provide much insight into the
physics of the system such as e.g. the large-scale behaviour of the evolving
particle density as a function of space and time, phase transitions,
and other phenomena that characterize the macroscopic state of the
system in the thermodynamic $L\to\infty$. The reason for this failure
is twofold. 

First of all, the discrete continuity equation cannot, in general,
be solved exactly. This can be seen by noting that generically the expectation of the instantaneous current involves correlations of higher
order. Taking further time-derivatives of these correlations, one is usually faced with the infinite BBGKY hierarchy of equations \cite{Spoh91}.
Thus usually it is not possible to obtain a closed system of equations for a finite
set of observables. 

The second reason arises from the fact that
in general the expectation of a local quantity, which is an average
over initial states and over infinitely many realizations of the stochastic dynamics, does not 
automatically
describe the typical behaviour of the particle system observed on 
coarse-grained scale in a {\it single} realization of the stochastic dynamics.
This latter problem, however, is what interests us: When observing, e.g.,
a piston filled with a gas, we do {\it not} wish to compute how the gas 
expands under heating on average in a very large number of experiments.
We want to compute how it behaves in any single
experiment, corresponding to a single realization of the underlying
effectively stochastic dynamics. This is possible because on macroscopic scale
the gas behaves deterministically always in the same way.  Hence averaging over realizations of the
process does not provide new information, studying the large-scale 
behaviour of a large number of atoms in a single realization is sufficient.

For a variety of stochastic interacting particle systems with one conserved
species of particles these two problems have been overcome mathematically 
rigorously by establishing that the large scale evolution of the {\it coarse-grained} 
particle density $\rho(x,t)$ is governed by the macroscopic continuity equation
\begin{equation}
\partial_t \rho(x,t)  + \partial_x j(x,t) = 0
\label{mce}
\end{equation}
where in spatially homogeneous systems the current is typically of the form 
\begin{equation}
j(x,t) = j^\ast(\rho(x,t)) - D(\rho(x,t)) \partial_x \rho(x,t).
\end{equation}
Here $j^\ast(\cdot)$ is the stationary current density relation which like the collective
diffusion coefficient $D(\cdot)$ depend only on the local density $\rho(x,t)$
\cite{Spoh91,Kipn99}. Thus the hydrodynamic equation \eqref{mce} is a closed 
(in general nonlinear) deterministic partial differential equation 
for the macroscopic density profile $\rho(x,t)$. 
The macroscopic space point $x$ represents a rescaled 
microscopically large region
of the underlying lattice on which the microscopic model is defined.
More precisely, this rescaling amounts to introducing a lattice constant 
$a$ and looking at the particle system in a region on the lattice of length 
$n a$ around some site. In the limit $a\to 0$ this lattice site and the
region around it then represent
the space point $x$. The macroscopic time is also large compared to
the microscopic time and usually taken with a rescaling $t \rightarrow t/a^{z}$ for some suitably chosen scaling exponent $z$ that depends on the
particle system under consideration.

The emergence of the macroscopic
continuity equation \eqref{mce} from the microscopic 
discrete continuity equation \eqref{dcegen} rests on two generic properties of 
stochastic interacting particle systems. The first is the law of large numbers which asserts that the
random number of particles in a large region (i.e., a large segment
as described above)
converges to a deterministic density $\rho(x,t)$ times the size of the
segment. The second reason
is local stationarity which means that the system is locally in the stationary
state at density $\rho(x,t)$ which results from rescaling the microscopic time $t$ to become very large. Since in one dimension stationary correlations
are generically short ranged \cite{Garr90}, the law of large numbers can indeed
by applied to describe the system locally (on large scale) by a density
$\rho(x,t)$ and the current $j(x,t)$ depends on $x$ and $t$
only through stationary expectations which can be expressed as 
functions of $\rho(x,t)$. 

Local stationarity also implies that the microscopic details
of the initial distribution are washed out and do not matter. Therefore
one expects that only the coarse-grained initial density $\rho_0(x):=\rho(x,0)$ determines the
coarse-grained density $\rho(x,t)$ at a later macroscopic time $t$.
Since the two properties of
the law of large numbers in a large region and local stationarity
are generic, the hydrodynamic limit \eqref{mce}  is expected to be valid
generically, even when there is no rigorous mathematical proof 
\cite{Spoh91}. Indeed, even when -- like in boundary-driven systems -- 
correlations are not short-ranged but sufficiently weak, the hydrodynamic 
description is still valid and can be proved rigorously for some classes of stochastic interacting particle systems
\cite{Eyin91,Baha12,Gonc19}. 

This picture is well-established for particle systems with short-range interactions. It cannot, however, be applied naively to stochastic interacting particle systems with long-interactions, for which exact results are scarce \cite{Ligg80,Andj03,Gonc18,Beli24} and where the notion
of 
locality of the current, i.e., its dependence as a function of the local density, becomes open to debate.
To address this issue and show that this locality may indeed get lost, we compute
in this work the large scale dynamics of a particular
stochastic interacting particle system with long-range interactions, viz., a symmetric exclusion process where the transition rates depend via a logarithmic pair potential not only on the occupation
of nearest neighbor sites, but on the entire configuration of the system.

As elaborated below, this process arises in similar form in a wide variety of contexts 
and it may thus be seen as 
paradigmatic. Using tools from condensed matter theory 
we show that at large space- and time scales
the expected local density appearing in the discrete continuity equation
\eqref{dcegen} is governed by a closed hydrodynamic equation of the form \eqref{mce}
where, however, the current $j(x,t)$ is {\it not} a local function
of the density $\rho(x,t)$, but a functional that via a Hilbert transform 
depends on the complete density profile at all points $y$ in space. This demonstrates that local stationarity remains crucial
also in the presence of long-range interactions, but with the
generalization that local variables are determined by the stationary
various everywhere instead of only locally.

This paper is organized as follows. In Sec. \ref{Sec:SDEP} we introduce the Symmetric Dyson Exclusion Process and point out some of its origins and connections to related problems. 
The stochastic dynamics is defined in terms of the intensity matrix
\cite{Ligg10,Schu01,Alca94}
whose off-diagonal elements are the transition rates and whose diagonal elements guarantee conservation of probability. In Sec.~\ref{Sec:hydro} we formulate the emergent macroscopic
description: starting from the exact microscopic expressions we propose
and motivate a \emph{non-local hydrodynamic equation}. We discuss its
equivalence to a local two-field (complex Hopf) hydrodynamic equation and its relation to other works.
Section~\ref{sec:numerics} tests these predictions against large-scale Monte-Carlo simulations.
We treat the ``melting’’ of single- and double-block initial states. We uncover the emergence of a space-time limit shape phenomenon that are characteristic to the model, Fig. ~\ref{fig:threeHoriz}, and derive explicit arctic curves, and find excellent
agreement with numerical data.

\section{Symmetric Dyson Exclusion Process
(SDEP)}
\label{Sec:SDEP}

We consider the symmetric Dyson exclusion process, denoted below by the 
acronym SDEP, which is an exclusion process with nearest 
neighbor jumps and a long-range interaction that arises in various
seemingly unrelated contexts. This process was first introduced as a 
model for steps of a vicinal surface in \cite{Spoh99} and later 
independently in \cite{Popk10,Schu15} by maximally conditioning the 
conventional Symmetric Simple Exclusion Process (SSEP) on a large number of particle jumps following the 
approach of \cite{Jack10,Chet15}. For the convenience of the reader, we provide a brief overview of this conditioning in the Appendix. The invariant measure of the SDEP, 
reviewed below, is described by the distribution of eigenvalues of random unitary 
matrices that perform Dyson's Brownian motion over the unitary group 
$U(N)$, thus making the SDEP a discrete interacting random walk version 
of Dyson's Brownian motion \cite{Dyso62}. It is also closely related to
classical point charges on a circle with a repulsive force arising from the 
two-dimensional Coulomb law. Endowed with Brownian motion, the 
dynamics of these point charges is intimately related to the 
Calogero–Sutherland model \cite{Calo69,Suth72} which has connections
to the Kadomtsev–Petviashvili equation \cite{Aban09,Kric94}.

Importantly, by conditioning the SSEP on an atypical activity, i.e. on an atypical number of particle jumps, the intensity matrix of the conditioned process is, 
up to an energy shift, the ground state transformation of the ferromagnetic 
Heisenberg spin-1/2 XXZ quantum chain which exhibits a phase transition 
from a phase-separated domain wall state \cite{Leco12}
to a hyperuniform conformally invariant phase \cite{Jack15a,Kare17}.
The phase transition occurs at a disorder point that corresponds to the unconditioned process, i.e., 
the conventional SSEP 
\cite{Schu01,Spit70,Ligg99}. When conditioned to maximal activity, the intensity matrix is the ground state transform of the Hamiltonian of the
spin-1/2 XX quantum chain (see Eq.~(\ref{GSTdef})), denoted below by $H^{{\rm XX}}$.


\subsection{Definition of the SDEP}

In the conventional SSEP defined
on a chain of $L$ sites with periodic boundary conditions,  each lattice site $k$ of a lattice of $L$ sites is occupied by at most particle. 
Particles in an ergodic sector with $N$ particles are labelled sequentially by integers $i\in\{1,\dots,N\}$. The $i$-th particle at
position $k_i$ attempts to jump after an exponential random time with parameter
$w$ to a neighboring site $k\pm 1$ with equal probability 1/2. If the selected target is empty then it
jumps, otherwise the jump attempt is rejected  and the particle stays at site $k_{i}$, thus respecting the on-site exclusion interaction. 
Below we label a configuration with particles
at sites $k_1,\dots,k_{N}$ by $\bfk$.

In the SDEP the interaction is long-ranged. The $i$-th particle at
is the average position $k_i$ attempts to jump to the neighboring site 
$k_i \pm 1$ with the rate
\begin{equation}
w_{N}^\pm(i) 
= \frac{w}{2} \prod\limits_{\substack{j=1\\j\neq i}}^{N}
\frac{\sin{\pi \frac{k_{j}-k_{i}\mp 1}{L}}}
{\sin{\pi\frac{k_{j}-k_{i}}{L}}}
\label{wipm}%
\end{equation}
which depends on the position of all other particles on the lattice.
The factor $w$ sets the time scale of the process.
For a fixed particle configuration $\bfk=(k_1,\dots,k_N)$, the quantity
$j_k^{\rm inst}(\bfk)$ denotes the microscopic instantaneous current across
the bond $(k,k+1)$, i.e. the net jump rate from $k$ to $k+1$ minus the net
jump rate from $k+1$ to $k$. It is given by
\begin{equation}
j^{{\rm inst}}_k
=
\frac{w}{2}
\sum_{i=1}^{N} \left(
\delta_{k_i,k}
\prod_{\stackrel{j=1}{j\neq i}}^{N} 
\frac{\sin{ \pi \frac{k_j-k_i-1}{L} } }
{\sin{ \pi \frac{k_j-k_i}{L} } } 
-  \delta_{k_i,k+1}
\prod_{\stackrel{j=1}{j\neq i}}^{N} 
\frac{\sin{ \pi \frac{k_j-k_i+1}{L} }}
{\sin{ \pi \frac{k_j-k_i}{L} } }
\right) .
\label{jkinst}
\end{equation}
The process is reversible w.r.t. the invariant measure \cite{Spoh99,Popk10}
\begin{equation}
\pi^\ast_{N}(\bfk) = \frac{1}{Z_{N} }
e^{-V_{N}(\bfk)}
\label{StationaryProbabilitiesSinVandermonde}
\end{equation}
with the long-range interaction potential 
\begin{equation}
V_{N}(\bfk)= - \sum\limits_{i=1}^{N-1}\sum\limits_{j=i+1}^{N} 
\ln{\left(\sin^2 \left( {\pi\frac{k_{j}-k_{i}}{L}}\right) \right)}
\label{U}
\end{equation}
and where $Z_N$ is a normalization factor \cite{Popk10}. 
Equation~(\ref{wipm}) is the natural nearest-neighbor reversible dynamics associated with the potential~(\ref{U}). Indeed, if $\bfk^{\,i,\pm}$ denotes the configuration obtained from $\bfk$ by replacing $k_i$ with $k_i\pm1$, then
\begin{equation}
w_N^\pm(i)=\frac{w}{2}\exp\!\left[-\frac12\bigl(V_N(\bfk^{\,i,\pm})-V_N(\bfk)\bigr)\right].
\end{equation}
Substituting Eq.~(\ref{U}) into this expression immediately reproduces the product form in Eq.~(\ref{wipm}). In other words, the rates are the square-root local-detailed-balance choice associated with the discrete Dyson log-gas weight $\pi_N^\ast(\bfk)\propto e^{-V_N(\bfk)}$, with the prefactor $w/2$ inherited from the bare symmetric nearest-neighbor dynamics. This also makes the reversibility with respect to $\pi_N^\ast$ transparent. As we recall below in Eq.~(\ref{GSTdef}), the same choice follows from the ground-state (Doob) transform of the XX chain.

The above-mentioned link with the eigenvalues $\exp{(2\pi i x_j)}$ of 
random unitary matrices 
that perform Dyson's Brownian motion over the unitary group 
$U(N)$ becomes apparent by the identification $k_j := x_j L$ of the 
parameters $x_j$ with the rescaled particle coordinates $k_j$. However, we stress that here the eigenvalues are restricted to the discrete set of $L^{\rm th}$ roots of unity.

\subsection{Some useful facts about the spin-1/2 XX quantum chain}

We label the lattice sites by integers $k\in\{-L/2+1,\dots, L/2\}$
modulo $L$. This is to facilitate taking the two-sided thermodynamic limit $L\to\infty$.
The spin-1/2 XX quantum chain is defined by the Hamiltonian
\cite{Lieb61}
\begin{equation}
H^{{\rm XX}} = - \frac{w}{4} \sum_{k = - L/2+1}^{L/2} \left(\sigma^x_{k} \sigma^x_{k+1} 
+ \sigma^y_{k} \sigma^y_{k+1} \right)
\end{equation}
and can be turned by a Jordan-Wigner transformation $\sigma_k^+ = \prod_{j=-L/1+1}^{k-1} (-1)^{c_j^\dagger c_j}  c_k^\dagger $ into a system
of spinless free fermions with fermionic creation and annihilation operators
$c^\dagger_k$, $c_k$ that satisfy the anticommutation relations
$\{c^\dagger_k, c_l\} = \delta_{k,l}$,
\begin{equation}
H^{{\rm XX}} = - \frac{w}{2} \sum_{k = - L/2+1}^{L/2} \left( c^\dagger_k c_{k+1} + c^\dagger_{k+1} c_{k} \right) .
\end{equation}
For self-containedness we recall some well-known properties of the 
XX-chain, see e.g. \cite{Lieb61,Niem67} for details. The $N$-particle states are defined by the ordered product
\begin{equation}
\ket{\bfk} :=
\prod_{n=1}^{\stackrel{N}{\to}} c^\dagger_{k_n} 
\ket{\emptyset} = c^\dagger_{k_1} c^\dagger_{k_2} 
\dots 
c^\dagger_{k_N}  \ket{\emptyset}
\label{bfkdef}
\end{equation}
where $\ket{\emptyset}$ is the vector representing the empty lattice
(or all spins up in spin language). Ordering means that for site indices $i\in\{1,\dots,N\}$ the particle
locations $k_i$ in $\bfk$ are from the subset $\Omega_{N}$ of 
$\{-L/2+1,\dots, L/2\}^N$ defined by
$k_j>k_i$ for $j>i$. 
The projection operator 
\begin{equation}
\hat{P}_N := \sum_{\bfk\in\Omega_{N}}
\ket{\bfk}\bra{\bfk}
\end{equation}
yields the Hamiltonian
\begin{equation}
H^{{\rm XX}}_N = H^{{\rm XX}} \hat{P}_N
\end{equation}
of the $N$-particle sector. The (normalized) ground state vector $\ket{0_N}$ is well known \cite{Niem67},
\begin{equation}
\ket{0_N} =
 \sum_{\bfk}
\Psi_{N}(\bfk) \ket{\bfk}
\label{GS}
\end{equation}
with the ground state wavefunction
\begin{equation} 
\Psi_{N}(\bfk) 
=  \frac{2^{N\choose 2}}{L^{N/2}} \prod_{1\le i<j\le N}
\sin{\left(\pi \frac{k_j-k_i}{L}\right)} .
\label{Psiff}
\end{equation}
Comparing Eq.~(\ref{Psiff}) with Eq.~(\ref{wipm}), one sees directly that for a nearest-neighbor move of particle $i$,
\begin{equation}
w_N^\pm(i)=\frac{w}{2}\,\frac{\Psi_N(\bfk^{\,i,\pm})}{\Psi_N(\bfk)},
\end{equation}
where $\bfk^{\,i,\pm}$ is obtained from $\bfk$ by replacing $k_i$ with $k_i\pm1$. This is precisely the off-diagonal structure generated by the ground-state transform in Eq.~(\ref{GSTdef}).
The corresponding ground state energy is~\cite{Niem67}
\begin{equation}
E_{N} = - w \frac{\sin{\left(\frac{\pi N}{L}\right)}}
{\sin{\left(\frac{\pi}{L}\right)}}
\label{J0N}
\end{equation}
and the spectral gap corresponding to an excitation of momentum $2\pi/L$ is
\begin{equation}
\Delta E_{N}  =2w \sin{\left(\frac{\pi N}{L}\right)}\sin{\left(\frac{\pi}{L}\right)} \simeq \frac{2\pi}{L}  \, w \sin{\left(\frac{\pi N}{L}\right)},
\label{SpectrumGap} 
\end{equation}
from which one reads off the sound velocity
\begin{equation}
    v_s = w \sin{(\pi\rho)}
\end{equation}
for density $\rho=N/L$.

\subsection{Intensity matrix of the SDEP}

For $N$ particles the intensity matrix $H_{N}$ of the SDEP is the Doob 
transform of the Hamiltonian of the XX-chain,
\begin{equation}
H_{N} := \hat{\Psi}_{N} \left(H_{N}^{{\rm XX}}  - E_{N} 
\right)\hat{\Psi}_{N}^{-1} ,
\label{GSTdef}
\end{equation}
where $\hat{\Psi}_{N} $ is the diagonal matrix
\begin{equation}
\hat{\Psi}_{N} := \sum_{\bfk\in\Omega_{N}} \Psi_{N}(\bfk) 
\ket{\bfk}\bra{\bfk} \, .
\end{equation}
Probability conservation is encoded in the $N$-particle summation
vector
\begin{equation}
\bra{s_{N}} = \sum_{\bfk\in\Omega_{N}} \bra{\bfk}
\label{sNdef}
\end{equation}
by the eigenvalue property $\bra{s_{N}} H_{N} = 0$.
The invariant measure is represented by the vector
\begin{equation}
\ket{\pi_N} := \sum_{\bfk\in\Omega_{N}} \Psi^2_{N}(\bfk) \ket{\bfk}
= \hat{\Psi}^2_{N} \ket{s_{N}}.
\end{equation}
Note that due to reversibility of the process the average value of the current in the steady state is
\begin{equation}
    \left< \sigma^+_k \sigma^-_{k+1} - \sigma^-_k \sigma^+_{k+1} \right> \, = \, 0
\end{equation}
while for the average value of the activity one finds from \cite{Niem67}
\begin{equation}
    \label{eq:activity_eq}
    \frac{w}{2}
    \left< \sigma^+_k \sigma^-_{k+1} + \sigma^-_k \sigma^+_{k+1} \right> \, = \, \frac{w}{L} \frac{\sin(\frac{\pi N}{L})}{\sin(\frac{\pi}{L})}
    \, \xrightarrow[L \rightarrow \infty]{} \, \frac{w}{\pi}  \sin(\pi \rho)
\end{equation}
where the thermodynamic limit is taken with a fixed density $ \rho = \frac{N}{L}$. This result for the average activity at equilibrium in the SDEP will play an important role below.

\section{Hydrodynamics}
\label{Sec:hydro}
We now turn to our main goal, which is to derive the coarse-grained hydrodynamics of the SDEP. In this Section we make a precise conjecture regarding the form of the hydrodynamic limit of the SDEP, which will be supported by numerical evidence presented in Sec.~\ref{sec:numerics}.

Note that, in the conventional SSEP, the spectral gap of the intensity matrix scales as $O(1/L^{2})$ corresponding to diffusive scaling with dynamical exponent $z=2$. This is of course consistent with the fact that the hydrodynamic limit of the SSEP is given by the heat equation~\cite{Ligg10}. In contrast, the spectral gap of the intensity matrix of the SDEP scales as $O(1/L)$, see Eq.~(\ref{SpectrumGap}), which leads to faster relaxation with dynamical exponent $z=1$. We thus expect that, upon coarse-graining under ballistic (Eulerian) scaling, the SDEP satisfies a hydrodynamic equation resulting from the continuity equation.
\begin{equation}
	\label{eq:continuity}
	\partial_t \rho(x,t) + \partial_x j(x,t) \, = \, 0
\end{equation}
with a current $j$ that turns out to be a functional of $\rho(x,t)$. Indeed, our main claim is that the exact form of the current is
\begin{equation}
	\label{eq:conjecture_j}
	j[\rho](x,t)  =  \frac{w}{\pi}\sin (\pi  \rho(x,t))  \sinh ( \pi \mathcal{H} \rho(x,t) ,
\end{equation}
where $\mathcal{H} f(x,t)$ is the Hilbert transform 
{w.r.t. the first argument of a function $f(x,t)$ that is periodic in $x$ with period $L$, defined as the principal value integral
\begin{equation}
	\label{eq:hilbert}
	\mathcal{H} f(x,t) =   {\rm p.v.}  \frac{1}{L}  \int_0^L  \frac{f(u,t) d u }{{\rm tan} \frac{\pi (x- u)  }{L} } \, .
\end{equation}
Since the Hilbert transform is not local, we see that the particle current (\ref{eq:conjecture_j}) is not local. 

We will provide compelling numerical evidence supporting this claim in Sec.~\ref{sec:numerics}. For now, let us motivate the form (\ref{eq:conjecture_j}) of the current, and discuss the relation between this claim and the existing literature.

\subsection{Heuristic derivation of the conjectured particle current (\ref{eq:conjecture_j})}
We now give a heuristic derivation of the conjectured macroscopic current
(\ref{eq:conjecture_j}) starting from the exact microscopic expression
(\ref{jkinst}). For a configuration with particle positions
$k_1,\dots,k_N$, let
\begin{equation}
n_k^{\rm inst}=\sum_{j=1}^N \delta_{k,k_j}\in\{0,1\}
\end{equation}
be the instantaneous occupation variable at site $k$, and define
\begin{equation}
S_k^{\rm inst}:=
\sum_{q\neq k,k+1} n_q^{\rm inst}
\log\!\left(
\frac{\sin\!\bigl(\pi \frac{q-k-1}{L}\bigr)}
{\sin\!\bigl(\pi \frac{q-k}{L}\bigr)}
\right).
\end{equation}
Then Eq.~(\ref{jkinst}) may be rewritten as
\begin{eqnarray}
\label{eq:jinst_guess}
\nonumber
j_k^{\rm inst}
&=&
\frac{w}{2}\,n_k^{\rm inst}(1-n_{k+1}^{\rm inst})\,e^{S_k^{\rm inst}}
-
\frac{w}{2}\,(1-n_k^{\rm inst})n_{k+1}^{\rm inst}\,e^{-S_k^{\rm inst}}.
\end{eqnarray}
The key point is that $S_k^{\rm inst}$ contains contributions from all other
particles. To separate the long-range and short-range parts, we introduce a
mesoscopic cutoff $1\ll \ell \ll L$ and split the sum defining $S_k^{\rm inst}$
into the regions $|q-k|\le \ell$ and $|q-k|>\ell$.
For the far part, one has
\begin{equation}
\log\!\left(
\frac{\sin\!\bigl(\pi \frac{q-k-1}{L}\bigr)}
{\sin\!\bigl(\pi \frac{q-k}{L}\bigr)}
\right)
=
-\frac{\pi}{L}\cot\!\left(\pi\frac{q-k}{L}\right)
+
O\!\left(\frac{1}{L^2\sin^2\!\bigl(\pi\frac{q-k}{L}\bigr)}\right),
\end{equation}
so that, for a smooth coarse-grained density profile $\rho$,
\begin{equation}
\sum_{|q-k|>\ell} n_q^{\rm inst}
\log\!\left(
\frac{\sin\!\bigl(\pi \frac{q-k-1}{L}\bigr)}
{\sin\!\bigl(\pi \frac{q-k}{L}\bigr)}
\right)
=
\pi\,\mathcal H\rho(k)
+
O\!\left(\frac{\ell}{L}\right)
+
O\!\left(\frac{1}{\ell}\right).
\end{equation}
Thus the contribution of particles farther than $\ell$ sites away self-averages
into the non-local field $\pi\,\mathcal H\rho(k)$, while the remaining
microscopic dependence is confined to a neighborhood of size $O(\ell)$ around
the bond $(k,k+1)$. Choosing, for instance, $\ell=L^{1/2}$ makes both error
terms vanish as $L\to\infty$.
This suggests that, after extracting the far-field factor
$e^{\pm\pi\mathcal H\rho}$, the remaining prefactor is local. Under local
equilibrium, its coarse-grained expectation should therefore depend only on the
local density $\rho(x,t)$. We are thus led to the Ansatz
\begin{equation}
j[\rho](x,t)
=
a_+(\rho(x,t))\,e^{\pi\mathcal H\rho(x,t)}
-
a_-(\rho(x,t))\,e^{-\pi\mathcal H\rho(x,t)},
\end{equation}
for some local functions $a_+$ and $a_-$.
By reflection symmetry, $a_+(\rho)=a_-(\rho)$. Writing
\begin{equation}
a_+(\rho)=a_-(\rho)=:\frac12\,a(\rho),
\end{equation}
the current takes the form
\begin{equation}
j[\rho](x,t)=a(\rho(x,t))\,\sinh\!\bigl(\pi\mathcal H\rho(x,t)\bigr).
\end{equation}
The same decomposition applied to the microscopic activity (obtained from
(\ref{eq:jinst_guess}) by replacing the minus sign between the two terms by a
plus sign) gives the coarse-grained activity
\begin{equation}
a[\rho](x,t)=a(\rho(x,t))\,\cosh\!\bigl(\pi\mathcal H\rho(x,t)\bigr).
\end{equation}
For a homogeneous equilibrium state of density $\rho$, one has
$\mathcal H\rho=0$, hence
\begin{equation}
a[\rho]=a(\rho).
\end{equation}
Therefore the prefactor $a(\rho)$ is naturally identified with the local
equilibrium activity $a_{\rm eq}(\rho)$, which yields
\begin{equation}
\begin{array}{rcl}
a[\rho](x,t) &=& a_{\rm eq}(\rho(x,t))\,
\cosh\!\bigl(\pi\mathcal H\rho(x,t)\bigr), \\[2mm]
j[\rho](x,t) &=& a_{\rm eq}(\rho(x,t))\,
\sinh\!\bigl(\pi\mathcal H\rho(x,t)\bigr).
\end{array}
\end{equation}
Finally, $a_{\rm eq}(\rho)$ is fixed by the exact lattice equilibrium result
(\ref{eq:activity_eq}), which leads to the conjectured current
(\ref{eq:conjecture_j}).

\subsection{Equivalence between non-local one-component hydrodynamics and local two-component hydrodynamics}

From now on we set the time scale $w=1$ to lighten the formulas.

It turns out that, because of the mathematical properties of the Hilbert transform, it is possible to reformulate the {\it non-local} one-component hydrodynamic equation
\begin{equation}
	\label{eq:1component}
	\partial_t \rho(x,t) + \partial_x [ \frac{1}{\pi} \sin (\pi  \rho(x,t))  \sinh ( \pi \mathcal{H} \rho(x,t) ) ] \, = \, 0 
\end{equation}
as a {\it local} two-component hydrodynamic equation, with two densities $\rho(x)$ and $\tilde{\rho}(x)$ that satisfy a system of coupled conservation laws:
\begin{equation}
	\label{eq:2component}
	\left\{ \begin{array}{ccl}
		\partial_t (\pi \rho(x,t)) + \partial_x [\frac{1}{\pi}  \sin (\pi  \rho(x,t))  \sinh ( \pi  \tilde{\rho} (x,t) ) ] & = & 0  \\
		\partial_t  (\pi \tilde{\rho}(x,t)) + \partial_x [ \frac{1}{\pi}  \cos (\pi  \rho(x,t))  \cosh ( \pi \tilde{\rho}(x,t) ) ] & = & 0 \,  .
	\end{array} \right.
\end{equation}
Eqs. (\ref{eq:1component}) and (\ref{eq:2component}) are equivalent as long as the initial conditions $\rho_0(x)$ and $\tilde{\rho}_0(x)$ at time $t=0$ satisfy
\begin{equation}
	\label{eq:initialhilbert}
	\tilde{\rho}_0(x) \, = \, \mathcal{H} \rho_0 (x) .
\end{equation}
The reason for this equivalence is as follows (see also the appendix of Ref.~\cite{bouchoule2020effect} for a similar calculation). The fundamental property of the Hilbert transform is that $f(x)$ and $\mathcal{H} f(x)$ are the real- and imaginary parts of the same analytic function in the upper half-plane. More specifically, for a periodic function $f(x)$ of period $L$, there  exists a function $F(z)$ defined for any ${\rm Im}\, z > 0$, that is analytic in the upper half-plane (including at infinity), with the same periodicity as $f$, i.e. $F(z+L) = F(z)$, and such that for all $x \in \mathbb{R}$
\begin{equation}
	F(x+ i \epsilon )  \, \underset{\epsilon \rightarrow 0^+}{=} \,  f(x) + i \mathcal{H} f(x) .
\end{equation}
To see why this implies the equivalence between (\ref{eq:1component}) and (\ref{eq:2component}), we observe that, if the initial condition (\ref{eq:initialhilbert}) is satisfied, then there is an analytic function $P_0(z)$ in the upper half-plane, that is periodic, and such that $P_0(x + i \varepsilon) \underset{\epsilon \rightarrow 0^+}{=}   \pi \rho_0(x) + i \pi \tilde{\rho}_0 (x)$.
Then we consider the partial differential equation
\begin{equation}
	\label{eq:hydroz}
	\partial_t P(z,t) + i \partial_z [ \cos ( P (z,t) ) ] \, = \, 0  ,
\end{equation}
with the initial condition $P(z,0) = P_0(z)$. Since the initial condition is analytic in the upper half-plane, $P(z,t)$ remains so at later times. Then we can identify the densities $\rho(x,t)$ and $\tilde{\rho}(x,t)$ with the real- and imaginary parts of $\frac{1}{\pi}P(z,t)$ along the real axis at any time,
\begin{equation}
	\rho (x,t) \, = \,  \lim_{\epsilon \rightarrow 0^+}  \frac{1}{\pi} {\rm Re} \, P(x + i \epsilon ,t)  , \qquad \tilde{\rho} (x,t) \, = \, \lim_{\epsilon \rightarrow 0^+}   \frac{1}{\pi}{\rm Im}\, P(x + i \epsilon ,t)  \quad \left(   = \, \mathcal{H} \rho(x,t) \right)
\end{equation}
and we see that Eqs.~(\ref{eq:1component}) and (\ref{eq:2component}) are both equivalent to (\ref{eq:hydroz}) by taking the real- and imaginary parts of that equation.

\subsection{Relation with Abanov's lattice free fermion imaginary-time hydrodynamics}
\label{eq:abanov}

The complex partial differential equation (\ref{eq:hydroz}) is not new. In the context of large-scale descriptions of systems of hard-core particles or spin chains in one dimension, it was discussed by Abanov in Ref.~\cite{abanov2006hydrodynamics}. In particular, as pointed out by Abanov, for the quantity $\zeta (z,t) =   \sin P(z,t)$  , it is equivalent to a complex Hopf (or inviscid Burgers) equation:
\begin{equation}
	\label{eq:complex_Hopf}
	\partial_t \zeta  - i \zeta \partial_z \zeta  \, = \, 0 .
\end{equation}
In this form, this equation has also appeared in various other contexts, as we briefly review below in Sec.~\ref{sec:other_works}. In Ref.~\cite{abanov2006hydrodynamics}, Abanov uses this equation to study the {\it emptiness formation probability} of the quantum XX spin chain. This is the probability that, upon measuring simultaneously the states of all spins in an interval in the ground state of the Hamiltonian $H^{\rm XX}$, one finds that they are all pointing down. This probability is found to decay as a Gaussian as a function of the interval length, and Abanov derives several exact results about this quantity using the complex Hopf equation. For this, he relies on the fundamental fact that the ground state can be obtained through an imaginary-time evolution:
\begin{eqnarray*}
	\left| {\rm ground \; state} \right>  \; \propto \; \lim_{t \rightarrow  +\infty}  e^{ - t H} \left| \psi_0 \right>   \;  \underset{t \rightarrow i t}{=} \; \lim_{ t \rightarrow  - i \infty}  e^{ - i t H} \left| \psi_0 \right>  ,
\end{eqnarray*}
where $\left| \psi_0 \right>$ is an arbitrary state with non-vanishing overlap with the ground state.

The {\it real-time} hydrodynamics of the quantum XX spin chain can be easily understood, as it reflects the dynamics of the underlying non-interacting fermions \cite{Antal1999}.
One way to proceed is to think of the non-interacting fermions semi-classically. On large scales, the state is described by a phase-space occupation function $W(x,p)$, where the momentum $p$ is defined modulo $2\pi$ because the model is considered on the lattice. Here $W$ should not be understood as a generic quantum Wigner quasi-probability, which need not be positive. Rather, in the zero-entropy semiclassical regime relevant here, $2\pi W(x,p)$ is the occupation of the phase-space cell $(x,p)$. By Pauli exclusion,
\begin{equation}
0\le 2\pi W(x,p)\le 1,
\end{equation}
or equivalently
\begin{equation}
0\le W(x,p)\le \frac{1}{2\pi}.
\end{equation}
The local particle density is obtained by integrating over momenta at a fixed position $x$,
\begin{equation}
	\label{eq:W_rho}
	\rho(x) \, = \, \int_{-\pi}^\pi dp \, W(x,p)  \quad \in \, [0,1] .
\end{equation}
The fermions move at their group velocity $v(p) = \sin (p)$, so $W(x,p)$ evolves in time according to the kinetic equation
\begin{equation}
	\label{eq:kinetic}
	\partial_t W + v(p) \partial_x W = 0 .
\end{equation}
The connection with the Hopf equation appears when one focuses on states that are locally equivalent to the ground state at some local density $\rho(x)$. Indeed, for such states, the phase-space occupation takes the form of a position-dependent Fermi sea, with an occupation function at position $x$ that is maximal in some interval of $p$ and vanishes in its complement: 
\begin{equation}
	W (x,p) \, = \, \left\{  \begin{array}{rcl}
		\frac{1}{2\pi} & {\rm if} &   p \in [ p_- (x), p_+ (x) ]  \\
		0 & {\rm if} &   p  \in [  p_+(x),  2\pi + p_- (x) ]	\, .
	\end{array} \right.
\end{equation}
Here $p$, $p_+$ and $p_-$ are all defined modulo $2\pi$. In such states the local density (\ref{eq:W_rho}) is
\begin{equation}
	\rho(x) \, = \,  \frac{ p_+(x) - p_-(x) }{2\pi} .
\end{equation}
 The kinetic equation (\ref{eq:kinetic}) boils down to a Hopf-like equation for the {\it Fermi points} $p_\pm(x,t)$,
\begin{equation}
	\label{eq:hydro_XX_realtime}
	\partial_t p_\pm  +  \sin ( p_\pm ) \partial_x p_\pm \, = \, 0  .
\end{equation}
The fact that the real-time dynamics of lattice free fermions initiated in states of zero entropy is given by Eq.~(\ref{eq:hydro_XX_realtime}) has been studied by many authors~\cite{Antal1999,Bettelheim2006,Bettelheim2012,Ruggiero2019,Scopa2021}. Abanov's main point is that questions related to the XX chain in {\it imaginary time} may then be tackled by considering the Wick rotation $t \rightarrow - i t$ of  Eq.~(\ref{eq:hydro_XX_realtime}). This is precisely the equation (\ref{eq:hydroz}) that we have encountered in the previous section, which is equivalent to the complex Hopf equation (\ref{eq:complex_Hopf}).

\subsection{Relation to other works}
\label{sec:other_works}

Imaginary-time hydrodynamics of free fermions has connections with many other problems. For instance, mutually avoiding directed polymers can be mapped to the worldlines of fermions in Euclidean time, a fact pointed out by de Gennes in the 1960s~\cite{deGennes1968}. The complex Hopf (or inviscid Burgers) equation (\ref{eq:complex_Hopf}) is also ubiquitous in problems of dimer coverings~\cite{Kenyon2007,Allegra2016,Gorin2021}, in certain random matrix problems such as the evaluation of the asymptotics of the Harish–Chandra–Itzykson–Zuber integral~\cite{Matytsin1994}, in studies of the emptiness formation probability~\cite{Korepin1993,Abanov2003,abanov2006hydrodynamics,Abanov2005,Stephan2014} and full counting statistics in spin chains~\cite{Pallister2025}, or in fluid two-dimensional fluid dynamics where it appears in connection with the Marangoni effect~\cite{Crowdy2021}.

More closely related to the SDEP, a series of recent works has focused on the continuous Dyson gas of Brownian particles with Coulomb repulsion on the continuum line
\cite{Andraus2016,dandekar2023,Dandekar2024,Krapivsky2025}, with a perspective that is similar to ours. In fact the SDEP can be viewed as the \emph{lattice} analogue of the continuous Dyson gas; the main structural
difference is the hard–core constraint $\rho\le 1$ that comes
automatically in a lattice exclusion process. For the continuous Dyson gas, the hydrodynamic equation of Refs.~\cite{Andraus2016,dandekar2023,Dandekar2024,Krapivsky2025} reads
\begin{equation}
    ({\rm continuous \; Dyson \; gas}) \qquad \quad  \partial_t \rho + \partial_x [\pi \rho \mathcal{H} \rho] \, = \, 0 .
\end{equation}
This equation can easily be recovered from our hydrodynamic equation for the SDEP (Eq.~(\ref{eq:1component})) by taking the low-density limit $\rho(x,t) \ll 1$.


\section{Density profile and limit shape for block initial states}
\label{sec:numerics}

In this section we provide numerical checks of the validity of the hydrodynamic equation (\ref{eq:1component}). We numerically simulate the microscopic dynamics of the SDEP using a kinetic Monte Carlo algorithm. For concreteness, we focus on two simple initial states: $N$ particles packed in a single block, and $N = N_1 + N_2$ particles packed into two separate blocks, see Fig.~\ref{fig:density2}.

\subsection{Density Profiles}

\begin{figure}[htbp]
  \centering
    \includegraphics[width=0.48\textwidth]{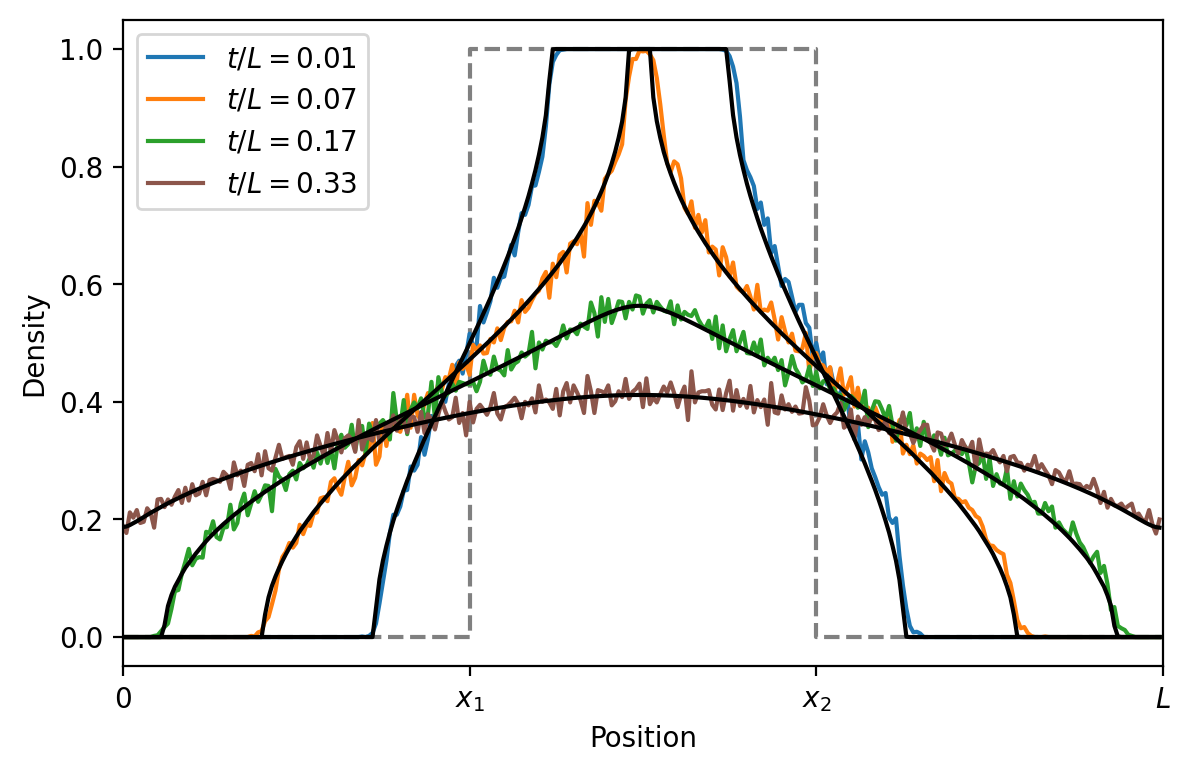} 
    \includegraphics[width=0.48\textwidth]{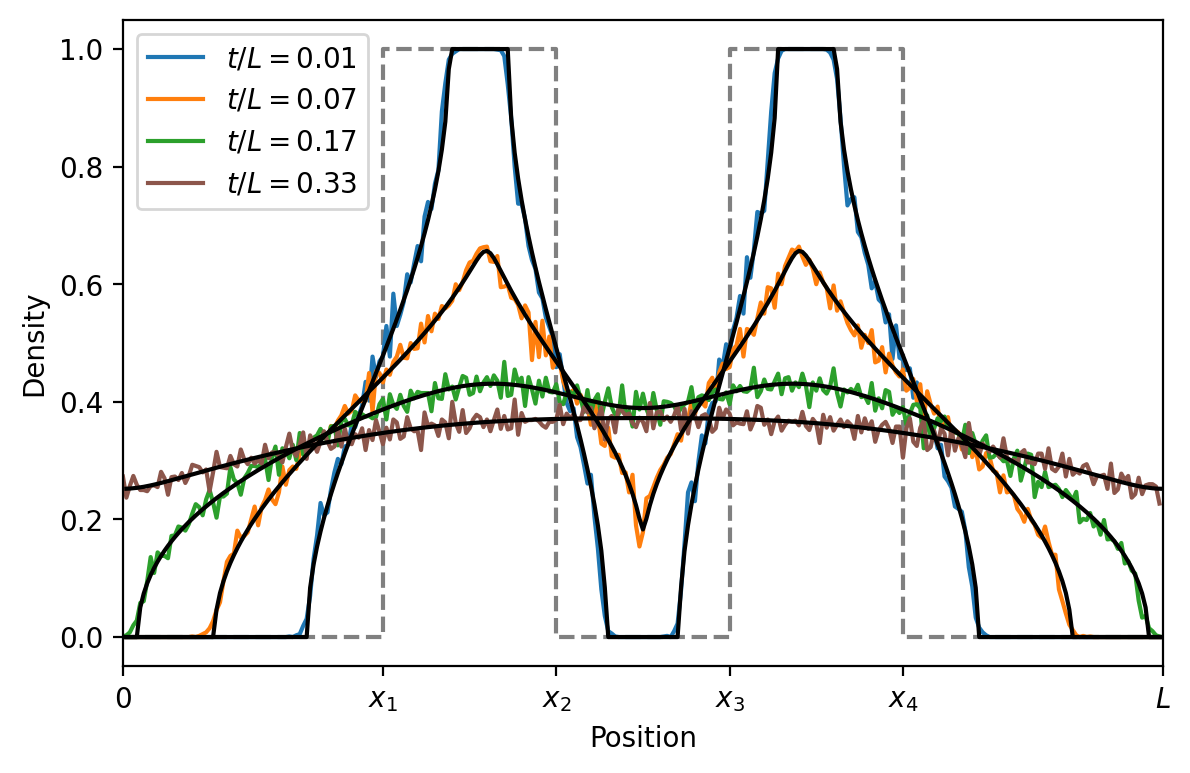} 

  \caption{Evolution of the density profile from a single block initial condition (left) and from a double-block initial condition (right); the dashed gray line shows the initial density profile. We compare the result of the stochastic simulation of the SDEP for $N=100$ particles on $L=300$ sites with periodic boundary conditions, to the hydrodynamic prediction (black continuous lines) obtained by solving equation (\ref{PP}) numerically. The profiles for the stochastic evolution are averaged over 1000 independent realizations.}
  \label{fig:density2}
\end{figure}

We start by comparing the density profiles at different times obtained from a direct stochastic simulation of the SDEP to the solution of the hydrodynamic equation (\ref{eq:1component}).
The Monte Carlo simulation is performed using a Gillespie-type algorithm, standard for continuous-time Markov processes. Starting from a configuration \(C_{t_i}\), the next configuration \(C_{t_{i+1}}\) is chosen from the set of accessible configurations with probability proportional to the transition rate from \(C_{t_i}\) to \(C_{t_{i+1}}\), after a waiting time \(t_{i+1}-t_i\) drawn from an exponential distribution with rate equal to the sum of outgoing transition rates from \(C_{t_i}\) to all accessible configurations.

To obtain the solution of the hydrodynamic equation (\ref{eq:1component}), we rewrite it in the form (see Eq.~(\ref{eq:hydroz}))
\begin{equation}\label{eq:cons}
    \partial_t P(x,t) - i \sin(P(x,t)) \partial_x P(x,t) = 0
\end{equation}
where $P(x,t) = \pi \rho(x,t) + i\pi \mathcal{H} \rho(x,t)$.
Similarly to the real inviscid Burgers equation, this equation admits solutions of the form
\begin{equation} \label{PP}
    P = P_0 (x + i t \sin(P)),
\end{equation}
where $P_0$ is an analytic function in the upper-half plane, determined by the initial condition at $t=0$,
\begin{equation}
{\rm Re}\,  P_0(x +i 0^+) = \pi \rho_0(x) .
\end{equation}
For the single block, corresponding to the initial condition $\rho_0(x) = 1$ if $x \in [x_1,x_2]$ and $\rho_0(x)=0$ otherwise, we take $P_0(z)$ as
\begin{equation}
    \label{eq:F1block}
 ({\rm single \; block}) \qquad P_0(z) 
 = i \, \ln\!\left[
      \frac{\sin\!\bigl(\frac{\pi}{L}(z-x_{1})\bigr)}
         {\sin\!\bigl(\frac{\pi}{L}(z-x_{2})\bigr)}
    \right],
\end{equation}
while for the two blocks $[x_1,x_2] \cup [x_3,x_4]$, where $\rho_0(x) = 1$ if $x \in [x_1,x_2] \cup [x_3,x_4]$ and $0$ otherwise, we have
\begin{equation}
    \label{eq:F2block}
 ({\rm two \; blocks}) \qquad P_0(z) = i \, \ln \left[  \frac{\sin\!\bigl(\tfrac{\pi}{L}(z-x_{1})\bigr)  \sin\!\bigl(\tfrac{\pi}{L}(z-x_{3})\bigr) }{ \sin\!\bigl(\tfrac{\pi}{L}(z-x_{2})\bigr)  \sin\!\bigl(\tfrac{\pi}{L}(z-x_{4})\bigr) }  \right] \, .
\end{equation}
Here `$\ln$' stands for the principal logarithm, with the branch cut along the negative real axis, i.e. $\ln z = \ln |z| + i \, \arg  z$.
The two expressions above are obtained by constructing an analytic, \(L\)-periodic
function in the upper half-plane whose boundary value satisfies
\({\rm Re}\,P_0(x+i0^+)=\pi\rho_0(x)\). For a single interval \([x_1,x_2]\), the
ratio
\begin{equation}
\frac{\sin\!\bigl(\frac{\pi}{L}(z-x_1)\bigr)}
     {\sin\!\bigl(\frac{\pi}{L}(z-x_2)\bigr)}
\end{equation}
has a phase jump of \(\pi\) precisely when \(x\in[x_1,x_2]\), and no jump
outside, so \(i\log(\cdot)\) has the required real part. For two disjoint
intervals, the corresponding contributions simply add, which gives the product
inside the logarithm.

To find the solution of Eq.~(\ref{PP}) for these two initial conditions, we use a fixed-point iteration method. We have observed that this method converges for all values of $(x,t)$ that we tested, so that one can numerically evaluate $P(x,t)$ easily. The density profile $\rho(x,t)$ is then obtained as the real part of $\frac{1}{\pi} P(x,t)$. The resulting curves for $\rho(x,t)$ are plotted in Fig.~\ref{fig:density2}, and we see that they perfectly match the data points obtained from the Monte Carlo simulation of the SDEP. This confirms the validity of the conjectured hydrodynamic equation (\ref{eq:1component}).

\subsection{Limit shapes in spacetime}

In Fig.~\ref{fig:density2} we see that the regions where the density is initially $0$ or $1$ do not immediately melt. Instead these two regions have sharp boundaries that persist up to some time. Thus, when we look at the trajectories of particles in space-time for the SDEP, one clearly sees the appearance of a {\it limit shape phenomenon}, see Fig.~\ref{fig:threeHoriz}. We observe the emergence of a sharp boundary between `frozen' regions where the particle density is exactly $0$ or $1$, and `fluctuating' regions where the density is strictly between $0$ and $1$. Such limit shape phenomena occur in various statistical physics problems, including crystal growth~\cite{Rottman1984,Nienhuis1984}, dimer coverings~\cite{Jockusch1998,Elkies1992a,Elkies1992b}, statistics of random Young tableaux~\cite{Logan1977}, the six-vertex model with domain-wall boundary conditions~\cite{Korepin2000,Colomo2008,Palamarchuk2010,Allegra2016}, see e.g.~\cite{Stephan2021,Kenyon2009} for introductions. The curve separating the frozen regions and fluctuating regions is known as the {\rm arctic curve}~\cite{Jockusch1998,Colomo2010a,Colomo2010b,Granet2019,diFrancesco2019}. The hydrodynamic equation (\ref{eq:1component}) can be exploited to make a prediction for the arctic curve for the above initial conditions, which can then be compared to the stochastic simulation of the SDEP, providing further numerical evidence for the validity of the hydrodynamic equation (\ref{eq:1component}). This is what we do next.

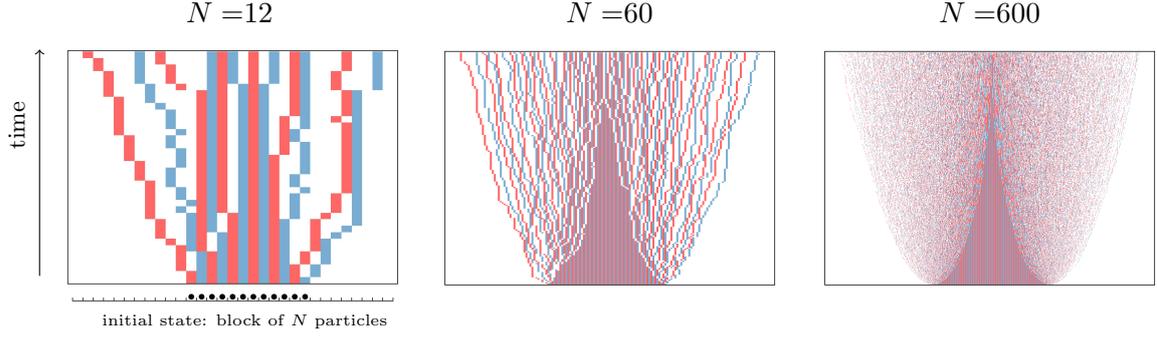
\begin{figure}[t]
    \centering
    
    \begin{tikzpicture}
        \draw (0,-0.152) node{\includegraphics[width=0.305\linewidth]{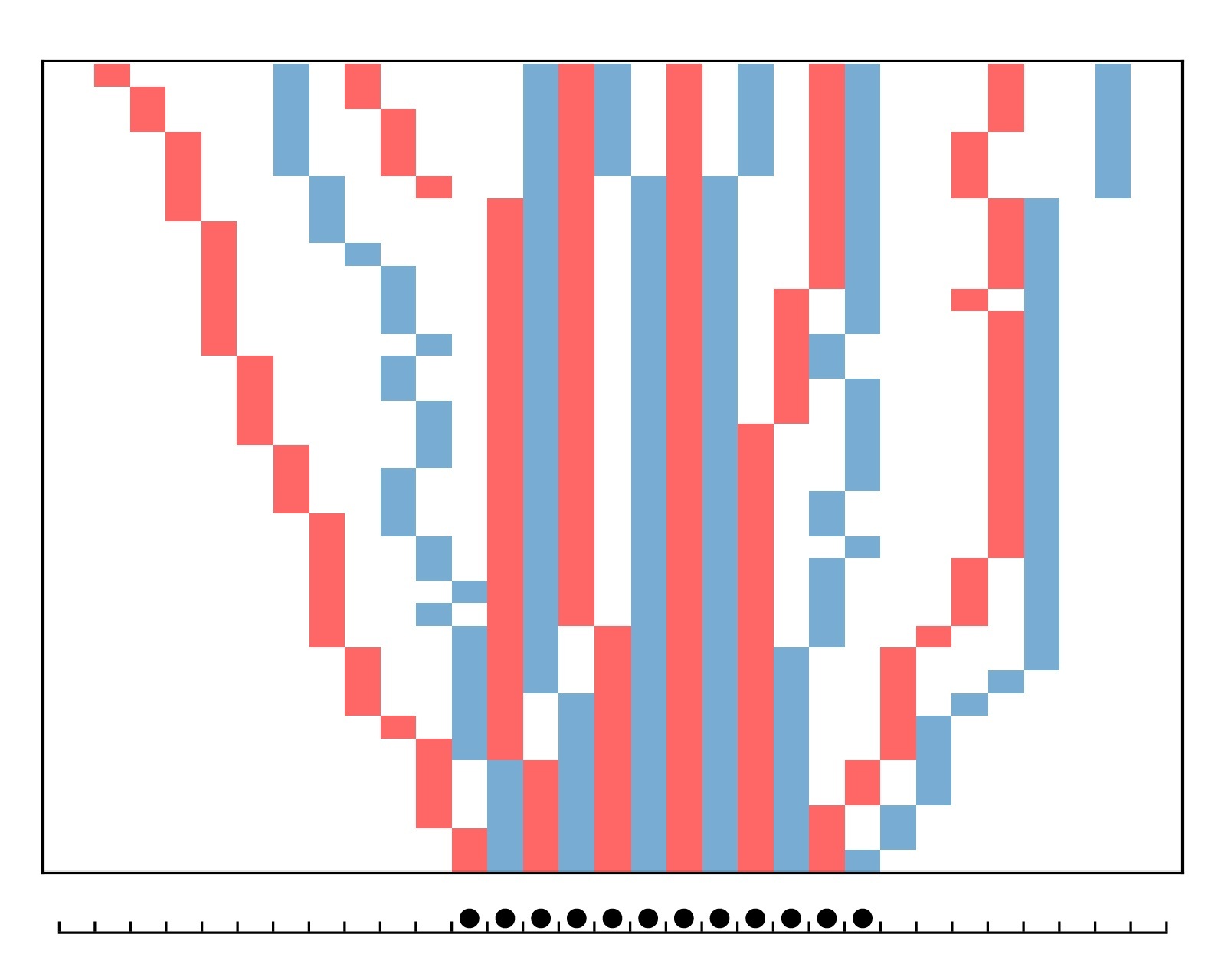}};
        \draw (5,0) node{\includegraphics[width=0.3\linewidth]{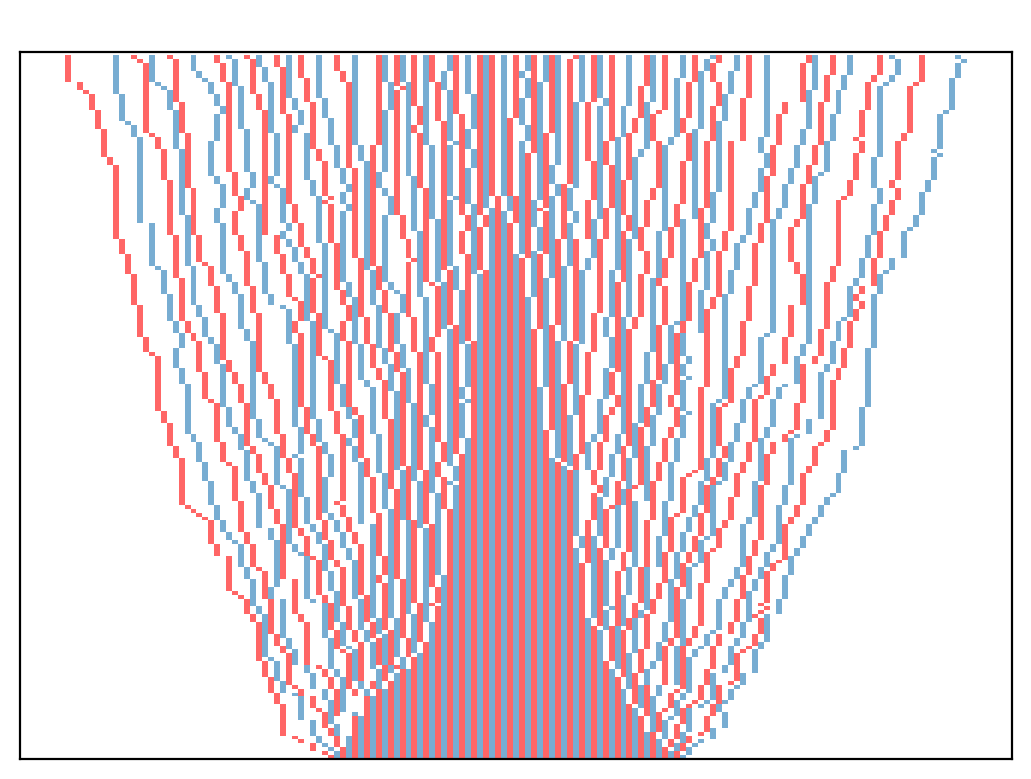}};
        \draw (10,0) node{\includegraphics[width=0.3\linewidth]{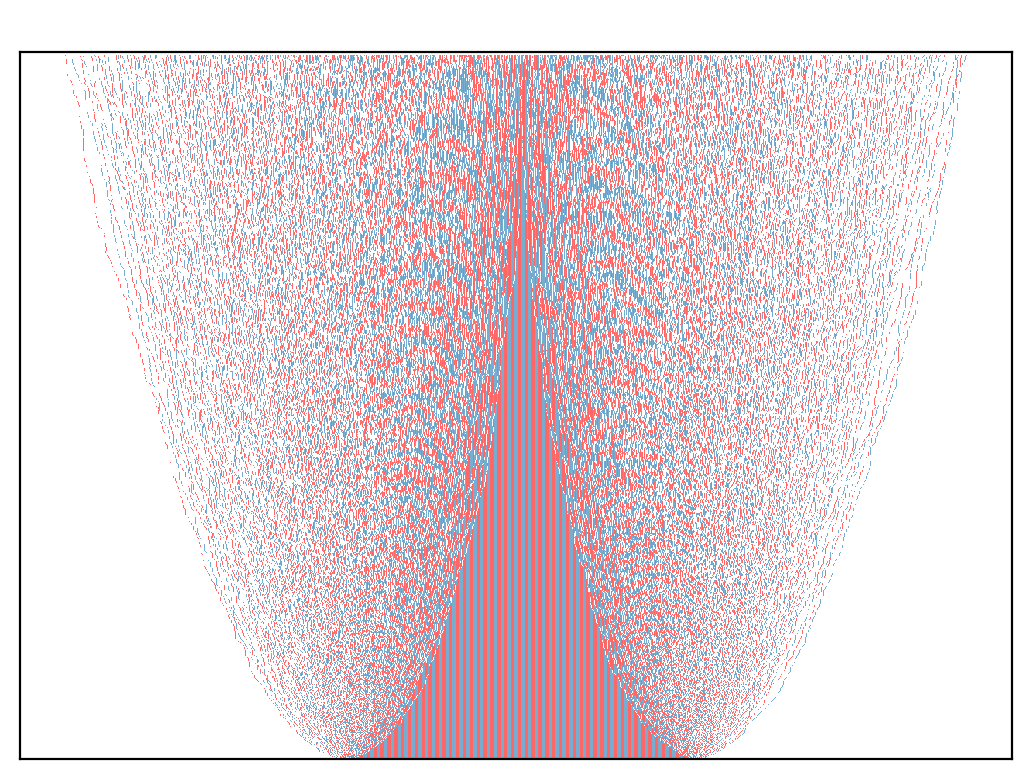}};
        \draw[->] (-2.5,-1.5) -- ++(0,3);
        \draw (-2.8,0.5) node[rotate=90]{{\footnotesize time}};
        \draw(0.2,-2.1) node{{\tiny initial state: block of $N$ particles}};
        \draw (0,2) node{$N=$12};
        \draw (5,2) node{$N=$60};
        \draw (10,2) node{$N=$600};
    \end{tikzpicture}

    \caption{Emergence of a limit-shape phenomenon as the system size grows:  the particle trajectories are plotted in spacetime, starting from an initial configuration consisting of a single block of $N$ particles on the infinite line. The initial configuration is shown below the first figure on the left. The vertical axis denotes time. Alternating colours are used solely to distinguish neighbouring trajectories. From left to right: $N=$12, 60, and 600 particles, respectively. We clearly see the emergence, as $N$ increases, of a sharp boundary between `frozen' regions where the density is $0$ or $1$, and `fluctuating' regions where the density is strictly between $0$ and $1$.  }
    \label{fig:threeHoriz}
\end{figure}

\subsubsection{Arctic curve for melting of a single block.}

To simplify the formulas and the determination of the arctic curve, from now on we take the thermodynamic limit $L \rightarrow \infty$. In this limit and 
for a finite number of particles $N$, the hopping rates Eq.~(\ref{wipm}) reduces to:
\begin{equation}
w_{N}^\pm(i) 
= \frac{1}{2} \prod\limits_{\substack{j=1\\j\neq i}}^{N}
\left[1 \mp
\frac{1}
{k_{j}-k_{i}}\right]
\end{equation}.
We focus on a single block initial state, with all sites $-\frac{N}{2}+1 , -\frac{N}{2}+2, \dots, \frac{N}{2}$ filled with $N \gg 1$ particles. Here we assume that $N$ is even, but this is not essential. It is convenient to introduce the rescaled position and time
\begin{equation}
    \label{eq:scaling}
    \xi = \frac{x}{N} \qquad \tau = \frac{t}{N} ,
\end{equation}
so that the initial hydrodynamic density profile is ($N \gg 1$)
\begin{equation}
\rho(\xi,\tau=0) =
\begin{cases}
1, & -\frac{1}{2} \leq \xi \leq \frac{1}{2}, \\
0, & \text{otherwise} .
\end{cases}
\end{equation}
As we have seen in the previous paragraph, the hydrodynamic profile at later times is given by $\rho(\xi, \tau) = \frac{1}{\pi} {\rm Re} \, P (\xi, \tau)$, where $P (\xi, \tau)$ is the solution of (see Eq.~(\ref{PP}))
\begin{equation} \label{PP1}
    P = P_0 (\xi + i \tau \sin(P)),
\end{equation}
where now $P_0(z)$ is given by the $L \rightarrow \infty$ limit of formula (\ref{eq:F1block}), 
\begin{equation}
P_0(z) = i \ln \left( \frac{z + \frac{1}{2}}{z - \frac{1}{2}} \right).
\end{equation}
Let $Z := e^{i P}$. Then we can rewrite equation \eqref{PP1} as
\begin{equation}
 Z = \frac{\tau\left(Z-\frac{1}{Z}\right) + 2 \xi - 1}{\tau \left(Z-\frac{1}{Z}\right) + 2\xi + 1},
 \label{ZZ}
\end{equation}
which is equivalent to a third-order polynomial equation,
\begin{equation}
   (Z - 1) ( - \tau  Z^2 + \tau - 2\xi Z - Z ) - 2Z = 0.
   \label{pol1}
\end{equation} 

\begin{figure}[!htbp]
  \centering
{\includegraphics[width=0.49\textwidth]{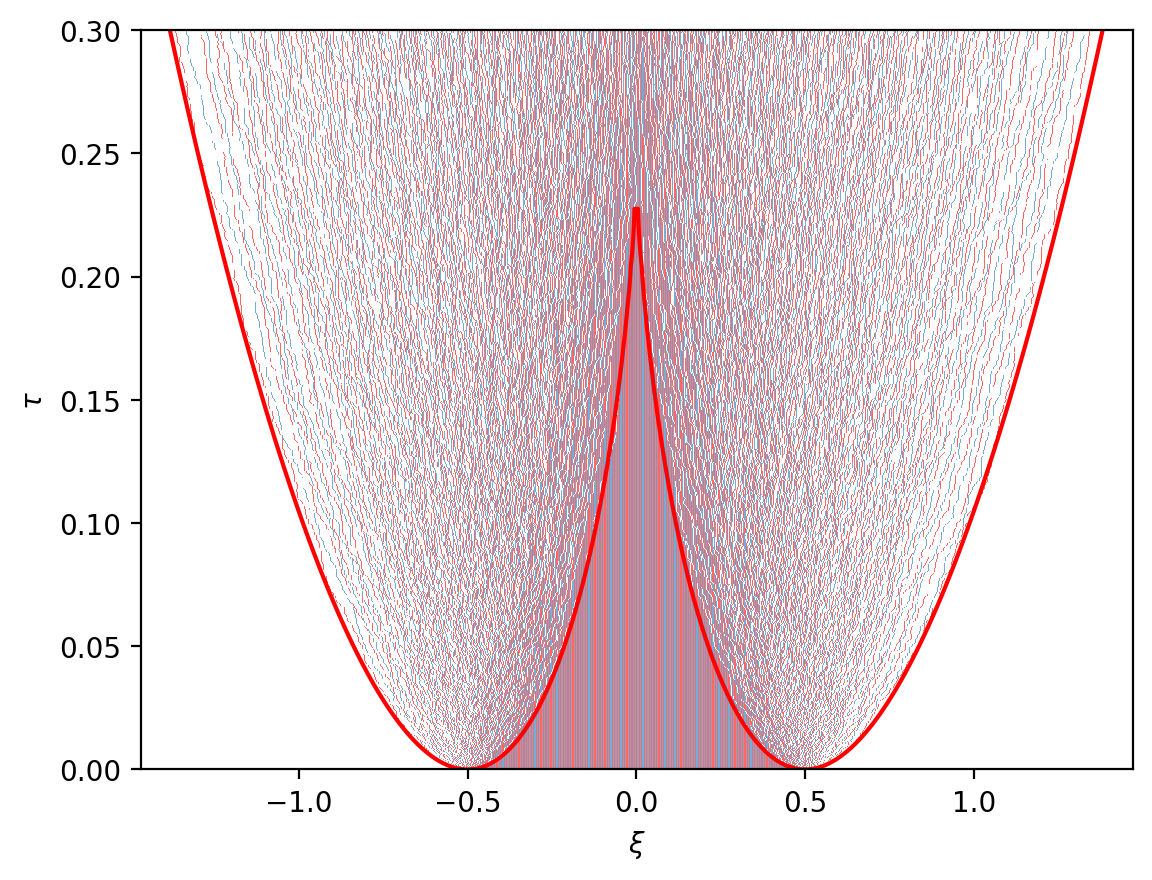}}\hfill
{\includegraphics[width=0.49\textwidth]{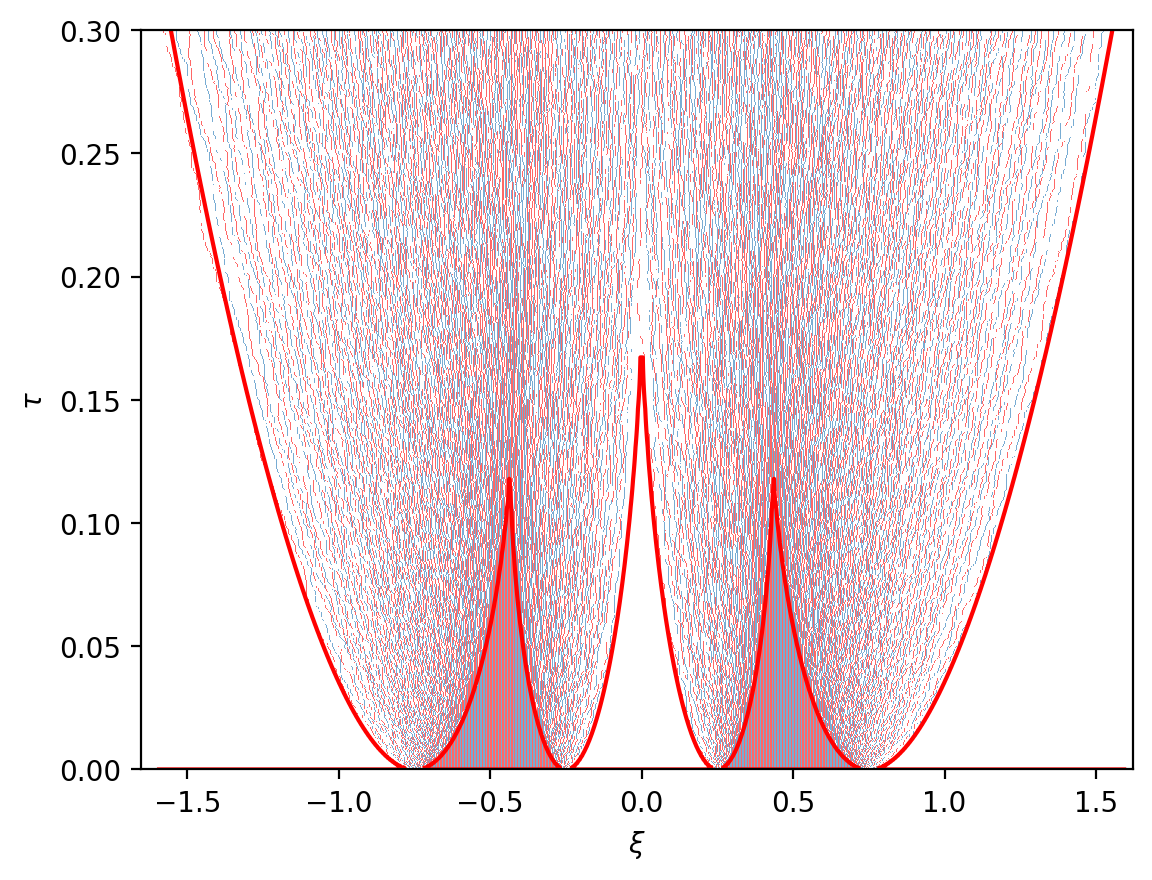}}
 \caption{The Arctic curve delineates the frozen regions, where the density is one inside and zero outside. Numerical simulation of particle trajectories from an initial configuration of a single block of $N=$ 400 particles centered at the origin (left) and two blocks each containing $N/2=$ 200 particles (right)
    }
\label{fig:mealting}
\end{figure}
Here and below, for a polynomial \(P(Z)=a_n Z^n+\cdots+a_0\) of degree \(n\), we denote by
\begin{equation}
\Delta(P)=(-1)^{n(n-1)/2} a_n^{-1}\,\mathrm{Res}(P,P')
\end{equation}
its discriminant, where \(\mathrm{Res}(P,P')\) is the resultant of \(P\) and its derivative \(P'\).
Equivalently, \(\Delta(P)=0\) if and only if \(P\) and \(P'\) have a common root, i.e. if and only if \(P\) has a repeated root.
The resultant can be computed directly from the coefficients of \(P\), for instance using the Euclidean algorithm.
For a real cubic, \(\Delta>0\) corresponds to three distinct real roots, whereas \(\Delta<0\) corresponds to one real root and one complex-conjugate pair.

Since the polynomial in the l.h.s has real coefficients, it has at least one real solution and at most two complex solutions that are complex conjugate. Given that $Z = \exp(i \pi \rho - \pi \tilde{\rho})$, we see that a density $\rho$ that is strictly between $0$ and $1$ corresponds to having a complex solution with $ \operatorname{Im}(Z) > 0$. This requires the discriminant of the above polynomial to satisfy $ \Delta < 0$. Accordingly, we should have
\begin{equation}
    \begin{array}{clc}
        \Delta > 0 & : & {\rm frozen \; region, \quad i.e. \;  } \rho = 0 \; {\rm or} \; \rho = 1 \\
        \Delta < 0 & : & {\rm fluctuating \; region, \quad i.e. \;  } 0 <\rho < 1 .
    \end{array}
\end{equation}
To find the `arctic curve' separating the two regions, we must solve the equation $\Delta = 0$. The discriminant of the polynomial (\ref{pol1}) is
\begin{equation}
\Delta = -64\tau^3 + 16\tau^2\xi^2 + 48\tau^2 - 80 \tau \xi^2 - 12\tau + 16\xi^4 - 8\xi^2 + 1 ,
\end{equation}
and by solving $\Delta = 0$ we find the arctic curve plotted in Fig.~\ref{fig:mealting} (left). The agreement with the emergent shape visible from the particle trajectories in the SDEP is compelling.

We note that at large $\tau$ the curve $(\xi, \tau)$ defined by $\Delta = 0$ is asymptotically the parabola $\xi(\tau) \simeq 2 \sqrt{\tau}$. Furthermore, one sees that the asymptotic density profile is a semi-circle that expands diffusively,
\begin{equation}
    \label{eq:semicircle}
    \rho(\xi,\tau) \underset{\xi, \tau \gg 1}{\simeq} \frac{1}{2\pi \tau} \sqrt{4 \tau - \xi^2} .
\end{equation}
To show this, notice first that in the limit of large $\tau$ and $\xi$  Eq.~(\ref{ZZ}) reduces to $Z \approx 1$. This suggests to perform an expansion $Z=1 + i P - P^2/2 + O(P^3)$. Inserting into Eq.~(\ref{pol1})
and keeping up to second order, we get:
\begin{equation}
2 + i (3 + 2 \xi) P - (5/2 + 2 \tau + 3 \xi) P^2 = O(P^3) .
\end{equation}
In order for this equation to have solutions, we need $P = O(\tau^{-1/2})$. Set $P = \frac{U}{\sqrt{\tau}}$ and $\xi = 2 \gamma \sqrt{\tau}$, with $\gamma \in (-1,1)$.
Keeping only the leading order in $\tau$ of the previous equation, we get $U^2-2 i \gamma U - 1  = O(\tau^{-1/2})$. Since ${\rm Re}(U) \geq 0$, we select the solution: $U = -i\gamma + \sqrt{1 - \gamma^2}$, which yields:
\begin{equation}
    P = \frac{-i\gamma}{\sqrt{\tau}} + \frac{1}{\sqrt{\tau}}  \sqrt{1 - \gamma^2} + O(\frac{1}{\tau}), \quad \gamma = \frac{\xi}{2 \sqrt{\tau}} .
\end{equation}
Taking the real part of both sides, we recover Eq.~(\ref{eq:semicircle}).

We note that the expanding semi-circle~(\ref{eq:semicircle}) is also found in the continuous Dyson gas~\cite{Krapivsky2025}. The reason why we recover the continuous gas behavior here is clear: by letting a block expand along the infinite line ($L\rightarrow \infty$), the density ultimately becomes very low, so our lattice gas becomes equivalent to a continuous gas, as discussed in  Sec.~\ref{sec:other_works}.


\subsubsection{Arctic curve for two blocks.}
Similar computations as in the previous paragraph can be done for the initial state with two blocks. Again, for simplicity we focus on the thermodynamic limit $L \rightarrow \infty$, and for concreteness we consider the following two initial blocks of particles occupying the sites $-\frac{3 N-2}{4}, -\frac{3 N-2}{4} +1 , \dots, -\frac{N+2}{4}$ and $\frac{N+2}{4}, \frac{N+2}{4}+1, \dots, \frac{3 N-2}{4}$ respectively. Here we are assuming that $N$ is of the form $N = 4p +2$ for some integer $p$.

We again use the scaling (\ref{eq:scaling}) for position and time, then the initial hydrodynamic density is
\begin{equation}
\rho(\xi) =
\begin{cases}
1, & \frac{1}{4} \leq |\xi| \leq \frac{3}{4}, \\
0, & \text{otherwise}.
\end{cases}
\end{equation}
To find the hydrodynamic density at later times, we again need to solve Eq.~(\ref{PP1}), but this time with 
\begin{equation}
P_0(z) = i \ln \left( \frac{z - \frac{1}{4}}{z - \frac{3}{4}}  \, \frac{z + \frac{3}{4}}{z + \frac{1}{4}} \right)
,
\end{equation}
which is the $L \rightarrow \infty$ limit of Eq.~(\ref{eq:F2block}).

Like in the single block case, Eq.~(\ref{PP1}) turns out to be equivalent to a polynomial equation for the variable $Z := e^{i P}$,
\begin{equation}\label{pol2}
4 \tau^2 (Z+1)^2 (Z-1)^3+4\tau Z \left(Z^2-1\right) (4 \xi (Z-1)+Z+1)+Z^2 (8 \xi (2 \xi (Z-1)+Z+1)-3 Z+3) = 0 .
\end{equation}
Then the arctic curve that separates the `frozen' region of spacetime ---where the density is $0$ or $1$--- from the `fluctuating' region ---where $0<\rho<1$--- is the set of points $(\xi,t)$ where the discriminant of this polynomial vanishes.
The discriminant (not shown here due to its length) can be obtained with the aid of symbolic computation software. The resulting equation is then solved numerically, yielding the curve shown in Fig.~\ref{fig:mealting} (right). Again, we see that the agreement with the stochastic simulation of the SDEP is compelling.
Performing a leading-order balance analysis on the discriminant yields an asymptotic arctic curve identical to the single-block case: $\xi = 2 \sqrt{t} + O(1)$

\section{Conclusion}

We have shown that the widely held expectation that reversible interacting particle systems generically display deterministic local hydrodynamics under \emph{diffusive} scaling
fails for the reversible symmetric Dyson exclusion process (SDEP) - a lattice gas with long-range interactions obtained by conditioning the SSEP on maximal activity and the discrete analogue of the Dyson log-gas. It possesses  genuinely non-local hydrodynamics under \emph{Eulerian} scaling whose low-density limit reproduces the continuum Dyson equation, yet exhibits intriguing limit-shape (arctic-curve) phenomena that spread diffusively 
at large macroscopic space-time scales and which are absent in the continuous case. 

Several open directions now suggest themselves. First, uncovering a genuinely “non-local Kardar-Parisi-Zhang (KPZ)’’ regime would be particularly exciting. Introducing a weak asymmetry or external noise is a natural route, yet it remains unclear which—if any—perturbations actually generate KPZ-type scaling. A link between free fermions 
to which the SDEP maps and  the KPZ universality class has been pointed out recently \cite{Imam23}. At the PDE level, a Hilbert-kernel variant of the KPZ equation has already been proposed earlier \cite{kechagia2001nonlocal}; clarifying its connection to the SDEP is an open problem. Second, applying macroscopic-fluctuation theory, as recently done for the continuum Dyson \cite{Dandekar2024} gas, should reveal how the hard-core constraint alters current large deviations; the long-range MFT developed in \cite{bernardin2025macroscopic} offers a natural starting point. Third, coupling the SDEP to particle reservoirs could help us explore basic questions such as how the boundary-induced phase transitions that are well understood for short range interactions \cite{hager2001minimal} would change in the presence of long range interactions. A few promising attempts are already present in the literature \cite{boccagna2020stationary,mourragui2014large}. Fourth, a full exploitation of the XX-chain mapping promises refined analytic control. This direction is currently being developed in ongoing work. Finally, conditioning the SSEP on finite (rather than maximal) activity, which corresponds to a Doob transform of the XXZ generator, remains a challenging but exciting avenue.

\section*{Acknowledgments}

\begingroup\setlength{\emergencystretch}{2em}

We thank Kirone Mallick for useful discussions. This work is supported by
ANR-PRME Uniopen (project ANR-22-CE30-0004-01),
FCT (Portugal) through project UIDB/04459/2020
(\href{https://doi.org/10.54499/UIDB/04459/2020}{\nolinkurl{10.54499/UIDB/04459/2020}})
and UIDP/04459/2020
(\href{https://doi.org/10.54499/UIDP/04459/2020}{\nolinkurl{10.54499/UIDP/04459/2020}}),
and by the FCT Grants 2020.03953.CEECIND/CP1587/CT0013
(\href{https://doi.org/10.54499/2020.03953.CEECIND/CP1587/CT0013}{\nolinkurl{10.54499/2020.03953.CEECIND/CP1587/CT0013}})
and 2022.09232.PTDC
(\href{https://doi.org/10.54499/2022.09232.PTDC}{\nolinkurl{10.54499/2022.09232.PTDC}}).
We acknowledge resources from Mésocentre EXPLOR of the University of Lorraine
(project \texttt{2024CPMXX3457}).
\par\endgroup

\appendix

\section[\quad \quad \quad \quad: \; Conditioning a Markov process on large deviations]{Conditioning a continuous-time Markov process on large deviations}

We provide a brief overview of how to condition a continuous-time Markov jump process on an atypical value of a time-additive observable using an exponential tilt and the associated Doob transform. This conditioning involves multiple steps, that we will go through one by one.
For detailed reviews see ~\cite{Jack10,Chet15,harris2007fluctuation}.

\paragraph*{Set-up and convention:}
Let $(X_t)_{t\ge 0}$ be a continuous-time Markov chain on a finite state space $\Omega$ with generator defined by
an intensity matrix $M\in\mathbb{R}^{|\Omega|\times|\Omega|}$. We adopt the \emph{column} convention: for $i\neq j$
$$
M_{ij} > 0 \quad\text{is the rate of the jump } j\to i,\qquad
M_{ii} = -\sum_{k\neq i} M_{ki}.
$$
Thus $\sum_i M_{ij}=0$ for each $j$, and the forward master equation is $\dot{\mathbf P}(t) = M\,\mathbf P(t)$ for the column probability vector $\mathbf P(t)$.

\paragraph*{Time-additive observable:}
Let
$
O \subset \{(i,j)\in\Omega\times\Omega: i\neq j\}²
$
be a specified set of ordered pairs of states (interpreted as jumps $i\to j$). Define the (extensive) counting process 
$$
J_t = \#\{\,u\in(0,t]: (X_{u^-},X_u)\in O \,\}.
$$
This variable $J_t$ increases by one each time the Markov chain performs the a transition from the set $O$. See \cite{harris2007fluctuation} for a detailed discussion. The associated \emph{time-averaged counting variable} is
$$
j_t := \frac{J_t}{t}.
$$
By ergodicity and the law of large numbers $j_t\to \bar{\jmath}$ almost surely as $t\to\infty$, where $\bar{\jmath}$ is the typical value determined by the stationary distribution of $M$ and the rates of jumps in $O$.
\paragraph*{Scaled cumulant generating function:}
To probe large fluctuations of $J_t$ (equivalently $j_t$), introduce the scaled cumulant generating function (SCGF)
\begin{equation}
\lambda(s) := \lim_{t\to\infty} \frac{1}{t}\,\log \mathbb{E}\!\bigl[e^{s J_t}\bigr].
\label{scgf}
\end{equation}
Provided this limit exists and is differentiable in a neighbourhood of interest, the random variables $j_t=J_t/t$ satisfy a large deviation principle
$$
\mathbb{P}(j_t \approx u) \asymp e^{-t\,I(u)}\qquad (t\to\infty),
$$
with rate function given by the Legendre--Fenchel transform
$$
I(u) = \sup_{s\in\mathbb{R}}\{\, s\,u - \lambda(s)\,\}.
$$
\paragraph{Tilted generator:}
Introduce the indicator function $f:\Omega\times\Omega\to\{0,1\}$ of the set $O$:
\begin{equation}\label{eq:f_indicator}
f(i,j) :=
\begin{cases}
1, & (i,j)\in O,\\
0, & \text{otherwise}.
\end{cases}
\end{equation}
We define the \emph{tilted generator} $M^{(s)}$:
\begin{equation}\label{eq:M_tilted}
M^{(s)}_{ij} = M_{ij}\,e^{s f(j,i)},\qquad i\neq j,
\end{equation}
Since $f(i,i)=0$, the diagonal is left unchanged:
$$
M^{(s)}_{ii} = M_{ii} = -\sum_{k\neq i} M_{ki}.
$$
The matrix $M^{(s)}$ is generally no longer a Markov generator (its column sums are not zero), but it is a real Metzler matrix (nonnegative off-diagonals).
For every state \(i\in\Omega\) define
\begin{equation}
  \psi_i(t;s)
  \;:=\;
  \mathbb{E}\!\Bigl[e^{\,sJ_t}\,\Big|\,X_t=i\Bigr],
  \label{eq:psi_hat_def}
\end{equation}
This is the generating function of the current \(J_t\) \emph{given that the
process ends in state \(i\) at time \(t\)}.

\smallskip
\noindent
Collecting the components produces a \emph{row} vector,
\begin{equation}
  \boldsymbol{\psi}(t;s)
    \;:=\;
    \bigl(\psi_i(t;s)\bigr)_{i\in\Omega}^{\top}
    \;\in\;\mathbb{R}^{|\Omega|}.
  \label{eq:psi_hat_vec}
\end{equation}
Because we have conditioned on the terminal state, this vector evolves
according to the \emph{backward} (adjoint) Kolmogorov equation:
\begin{equation}
  \frac{\mathrm d}{\mathrm dt}\,
  \boldsymbol{\psi}(t;s)
    \;=\;
    -\,\widehat{\boldsymbol{\psi}}(t;s)\,M^{(s)},
  \qquad
  \boldsymbol{\psi}(0;s)=\mathbf 1^{\!\top},
  \label{eq:psi_hat_ode}
\end{equation}
where $\mathbf 1$ is an all-ones column vector. The terminal condition reflects the fact that at
time \(t=0\) (when no time has elapsed) the observable \(J_0\) is zero
and the chain must be in its realised final state, so the conditional
expectation equals \(1\).

\smallskip
\noindent
Solving \eqref{eq:psi_hat_ode} gives:
\begin{equation}
 \boldsymbol{\psi}(t;s)
    \;=\;
    \mathbf 1^{\!\top}\,\exp\!\bigl(t\,M^{(s)}\bigr),
  \label{eq:psi_hat_solution}
\end{equation}
Starting from an initial state $P_0$, the unconditional generating function is given by:
$$
\mathbb{E}\bigl[e^{sJ_t}\bigr]=
\mathbf 1^{\!\top}e^{\,tM^{(s)}}\mathbf p_0,
$$
\paragraph{Spectral representation}
Assume the original chain is irreducible; then, for $s$ in a neighbourhood of 0 (and more generally whenever $M^{(s)}$ remains irreducible), the Perron--Frobenius theorem ensures that $M^{(s)}$ has a unique real, algebraically simple eigenvalue $\Lambda(s)$ with largest real part and strictly positive left and right eigenvectors $\mathbf l(s)$ and $\mathbf r(s)$:
\begin{align}
\mathbf l(s)^\top M^{(s)} &= \Lambda(s)\,\mathbf l(s)^\top,\\
M^{(s)} \mathbf r(s) &= \Lambda(s)\,\mathbf r(s),
\end{align}
which we normalise by $\mathbf l(s)^\top \mathbf r(s)=1$.
Then
$$
\mathbb{E}\bigl[e^{sJ_t}\bigr]
 = \bigl[\mathbf 1^\top \mathbf r(s)\bigr]\,
   \bigl[\mathbf l(s)^\top \mathbf p_0\bigr]\,
   e^{t\Lambda(s)}\,[1+o(1)],\qquad t\to\infty,
$$
and hence from \eqref{scgf} we identify
$$
\lambda(s) = \Lambda(s).
$$
Moreover $\lambda(0)=0$ and $\mathbf l(0)=\mathbf 1$.
\paragraph{Doob transform:}
Let
$$
\Delta^{(s)} := \mathrm{diag}\bigl(l_1(s),\dots,l_{|\Omega|}(s)\bigr).
$$
Define the \emph{Doob-transformed} generator
\begin{equation}\label{eq:Doob_transform}
M^{\mathrm{cond}}(s)
 := \Delta^{(s)} M^{(s)} (\Delta^{(s)})^{-1} - \lambda(s)\,I.
\end{equation}
Because $\mathbf l(s)^\top M^{(s)} = \lambda(s) \mathbf l(s)^\top$, it follows that the columns of $M^{\mathrm{cond}}(s)$ sum to zero and its off-diagonal entries are nonnegative; hence $M^{\mathrm{cond}}(s)$ is a valid CTMC generator (the “driven” or “conditioned” process).

Componentwise for $i\neq j$,
\begin{equation}\label{eq:Doob_offdiag}
M^{\mathrm{cond}}_{ij}(s)
= \frac{l_i(s)}{l_j(s)}\,M^{(s)}_{ij}
= \frac{l_i(s)}{l_j(s)}\,M_{ij}\,e^{s f(j,i)},
\end{equation}
and
\begin{equation}\label{eq:Doob_diag}
M^{\mathrm{cond}}_{jj}(s)
= M^{(s)}_{jj} - \lambda(s)
= M_{jj} - \lambda(s).
\end{equation}
\paragraph{Conditioned dynamics}
Let $\mathbf P(t)$ evolve under the original process: $\dot{\mathbf P}(t)=M\,\mathbf P(t)$. Define the reweighted measure
$$
\mathbf P^{\mathrm{cond}}(t)
= \frac{\Delta^{(s)} \mathbf P(t)}{\mathbf 1^\top \Delta^{(s)} \mathbf P(t)}.
$$
Differentiating and using \eqref{eq:Doob_transform} shows that
$$
\frac{d}{dt}\,\mathbf P^{\mathrm{cond}}(t)
= M^{\mathrm{cond}}(s)\,\mathbf P^{\mathrm{cond}}(t),
$$
so $\mathbf P^{\mathrm{cond}}$ evolves according to the Doob-transformed generator. (This equality holds exactly when the normalising scalar is tracked; see, e.g.,~\cite{Chet15,Jack10}.)
\paragraph{Invariant measure}
Let $\mathbf r(s)$ be the positive right Perron eigenvector of $M^{(s)}$:
$$
M^{(s)} \mathbf r(s) = \lambda(s)\, \mathbf r(s).
$$
Then
$$
M^{\mathrm{cond}}(s)\, \bigl(\Delta^{(s)} \mathbf r(s)\bigr)
= \Bigl(\Delta^{(s)}M^{(s)}(\Delta^{(s)})^{-1} - \lambda(s)I\Bigr) \Delta^{(s)} \mathbf r(s)
= \Delta^{(s)} M^{(s)} \mathbf r(s) - \lambda(s) \Delta^{(s)} \mathbf r(s)
= \mathbf 0,
$$
so $\Delta^{(s)} \mathbf r(s)$ is in the kernel of $M^{\mathrm{cond}}(s)$. Writing it componentwise we obtain the stationary (invariant) distribution
$$
\pi^{\mathrm{cond}}_i(s)
= \frac{l_i(s) r_i(s)}{\sum_k l_k(s) r_k(s)}.
$$
With our normalisation $\mathbf l(s)^\top \mathbf r(s)=1$, the denominator is 1 and $\pi^{\mathrm{cond}}_i(s)=l_i(s)r_i(s)$.
\paragraph{Mean current and derivative of the SCGF}
Differentiate the right-eigenvector relation $M^{(s)}\mathbf r(s)=\lambda(s)\,\mathbf r(s)$ with respect to $s$:
$$
M'^{(s)}\mathbf r(s) + M^{(s)}\mathbf r'(s)
= \lambda'(s)\mathbf r(s) + \lambda(s)\mathbf r'(s).
$$
Left-multiply by $\mathbf l(s)^\top$ and use $\mathbf l(s)^\top M^{(s)}=\lambda(s)\mathbf l(s)^\top$ and $\mathbf l(s)^\top \mathbf r(s)=1$ to get
$$
\lambda'(s) = \mathbf l(s)^\top M'^{(s)} \mathbf r(s).
$$
Because only off-diagonals of $M^{(s)}$ depend on $s$,
$$
M'^{(s)}_{ij} = f(j,i) M^{(s)}_{ij}, \qquad i\ne j;\qquad
M'^{(s)}_{ii}=0.
$$
Therefore
$$
\lambda'(s) = \sum_{i\ne j} f(j,i)\, l_i(s) M^{(s)}_{ij} r_j(s).
$$
Use \eqref{eq:Doob_offdiag} to substitute $M^{(s)}_{ij} = (l_j(s)/l_i(s)) M^{\mathrm{cond}}_{ij}(s)$:
$$
\lambda'(s) = \sum_{i\ne j} f(j,i)\, M^{\mathrm{cond}}_{ij}(s)\, l_j(s)r_j(s)
= \sum_{j} \pi^{\mathrm{cond}}_j(s) \sum_{i\ne j} M^{\mathrm{cond}}_{ij}(s)\, f(j,i).
$$
Recalling that under our column convention $M^{\mathrm{cond}}_{ij}(s)$ is the rate of $j\to i$ in the conditioned process, the inner sum is the \emph{expected instantaneous rate of jumps from $j$ into $O$}. Hence
$\lambda'(s)$ is the mean rate of $O$-jumps in the stationary conditioned dynamics.
By the ergodic theorem (law of large numbers) for the conditioned CTMC,
\begin{equation}
\lim_{t\to\infty} \frac{J_t}{t} = \lambda'(s)
\qquad \text{almost surely under }M^{\mathrm{cond}}(s)
\end{equation}
Thus the conjugate parameter $s$ continuously controls the typical current in the driven (conditioned) process.
\paragraph{Relation to micro-canonical conditioning:}
Let $j$ be a desired current density. Suppose the LDP rate function $I(u)$ is differentiable at $u=j$. Then the equation
$$
\lambda'(s)=j
$$
has a unique solution $s=s(j)$, and the long-time path measure of the original process conditioned on $J_t/t\approx j$ (microcanonical) is \emph{logarithmically equivalent} to the path measure of the Doob-transformed process $M^{\mathrm{cond}}(s(j))$ (canonical). That is, bulk observables computed in either ensemble agree in the $t\to\infty$ limit; see~\cite{Jack10,Chet15} and
\cite{Mont22, chetrite2013nonequilibrium} for details.

\bibliography{ref_sci}


\end{document}